\documentclass[a4paper,11pt]{article}
\usepackage[utf8]{inputenc}
\usepackage[english]{babel}
\usepackage[T1]{fontenc}
\usepackage{graphicx}
\usepackage{authblk}
\usepackage[title]{appendix}
\usepackage{caption}
\usepackage{csquotes}
\usepackage{color}
\usepackage{footmisc}
\usepackage{footnote}
\usepackage{epsfig}
\usepackage{amssymb}
\usepackage{latexsym}
\usepackage{amsmath}
\usepackage{pifont}
\usepackage{url}
\usepackage{listings}
\usepackage{xcolor}
\usepackage{hyperref}
\usepackage[colorinlistoftodos]{todonotes}

\newcommand{\vs}{{viewcounts }}
\newcommand{\vi}{{viewcount }}
\newcommand{\s}{{\sigma}}
\usepackage{rotating}
\usepackage[backend=biber,style=numeric-comp,sorting=none]{biblatex}
\usepackage{multirow}
\usepackage[normalem]{ulem}
\useunder{\uline}{\ul}{}
\DeclareUnicodeCharacter{0301}{\'e}

\usepackage{xr}
\makeatletter
\newcommand*{\addFileDependency}[1]{
  \typeout{(#1)}
  \@addtofilelist{#1}
  \IfFileExists{#1}{}{\typeout{No file #1.}}
}
\makeatother

\newcommand*{\maindoc}[1]{
    \externaldocument{#1}
    \addFileDependency{#1.tex}
    \addFileDependency{#1.aux}
}

\maindoc{appendices}

\bibliography{references}

\hypersetup{
colorlinks=true,
linkcolor=blue,
filecolor=magenta,
urlcolor=blue,
}

\author[1]{Milan Jovi\'c}
\author[2]{Lovro \v Subelj}
\author[3,4]{Tea Golob}
\author[3,4]{Matej Makarovi\v{c}}
\author[5,6]{Taha Yasseri}
\author[1]{Danijela Boberi\'c Krsti\' cev}
\author[4,1]{Srdjan Škrbi\'{c}}
\author[4,*]{Zoran Levnaji\'c}

\affil[1]{Department of Mathematics and Informatics, Faculty of Sciences, University of Novi Sad, Serbia} 
\affil[2]{Faculty of Computer and Information Science, University of Ljubljana, Slovenia}
\affil[3]{School of Advanced Social Studies, Nova Gorica, Slovenia}
\affil[4]{Faculty of information studies in Novo mesto, Slovenia}
\affil[5]{School of Sociology, University College Dublin, Ireland}
\affil[6]{Geary Institute for Public Policy, University College Dublin, Ireland}
\affil[*]{To whom the correspondence should be addressed: \texttt{zoran.levnajic@fis.unm.si}}

\colorlet{punct}{red!60!black}
\definecolor{background}{HTML}{EEEEEE}
\definecolor{delim}{RGB}{20,105,176}
\colorlet{numb}{magenta!60!black}

\lstdefinelanguage{json}{
    basicstyle=\normalfont\ttfamily,
    numbers=left,
    numberstyle=\scriptsize,
    stepnumber=1,
    numbersep=8pt,
    showstringspaces=false,
    breaklines=true,
    frame=lines,
    backgroundcolor=\color{background},
    literate=
     *{0}{{{\color{numb}0}}}{1}
      {1}{{{\color{numb}1}}}{1}
      {2}{{{\color{numb}2}}}{1}
      {3}{{{\color{numb}3}}}{1}
      {4}{{{\color{numb}4}}}{1}
      {5}{{{\color{numb}5}}}{1}
      {6}{{{\color{numb}6}}}{1}
      {7}{{{\color{numb}7}}}{1}
      {8}{{{\color{numb}8}}}{1}
      {9}{{{\color{numb}9}}}{1}
      {:}{{{\color{punct}{:}}}}{1}
      {,}{{{\color{punct}{,}}}}{1}
      {\{}{{{\color{delim}{\{}}}}{1}
      {\}}{{{\color{delim}{\}}}}}{1}
      {[}{{{\color{delim}{[}}}}{1}
      {]}{{{\color{delim}{]}}}}{1},
}

\begin{document}

\title{Terrorist attacks sharpen the binary perception of \\ 
``Us'' vs. ``Them''}

\date{}
\maketitle

\vspace*{-1.cm}
\begin{abstract} 
Terrorist attacks not only harm citizens but also shift their attention, which has long-lasting impacts on public opinion and government policies. Yet measuring the changes in public attention beyond media coverage has been methodologically challenging. Here we approach this problem by starting from Wikipedia's r\'epertoire of 5.8 million articles and a sample of 15 recent terrorist attacks. We deploy a complex exclusion procedure to identify topics and themes that consistently received a significant increase in attention due to these incidents. Examining their contents reveals a clear picture: terrorist attacks foster establishing a sharp boundary between ``Us'' (the target society) and ``Them'' (the terrorist as the enemy). In the midst of this, one seeks to construct identities of both sides. This triggers curiosity to learn more about ``Them'' and soul-search for a clearer understanding of ``Us''. This systematic analysis of public reactions to disruptive events could help mitigate their societal consequences.
\end{abstract}



\section{Introduction}

On the evening of 13 November 2015, three suicide bombers struck outside Stade de France in Paris. This was the beginning of the deadliest event in France since the Second World War, culminating in a mass shooting at the Bataclan theatre later that evening. The terrorist attacks claimed the lives of 137 people including the seven attackers'. Events such as these prompt debates infused with anger, accusations, and hatred, but also sometimes followed by self-reflection, introspection, and soul-searching \cite{Magdy2015,Strebel2017,Vasilopoulos2019}. Amid such turmoil, the public focus suddenly shifts to topics and issues that were marginal before the event and are not even necessarily related to it directly. Events in Paris that night sparked curiosity about Islam and its history, Western colonial legacy, and tourist spots in the Middle East, but also triggered discussions about West and East, Christianity and Islam, and ``Us'' and ``Them'' \cite{Strebel2017,Said1978,Rane2014,Kearns2019}. This confrontation has a regrettably violent history, which includes notable incidents such as the Twin Towers attack in 2001 and the London Underground attack in 2005. \\[0.2cm]
Here we ask: Which topics do come to the forefront of the public discourse in the wake of a terrorist attack? Are there any common patterns across various attacks? How long does it take for public attention to return to routine concerns? Past research in this direction has mostly focused on the psychology and politics behind the terrorist attacks and their portrayal in the media \cite{Slone2000,Pfefferbaum2014,Iyer2014,Conway2017,Hopwood2017,Monfort2017,thompson2019media,Rime2019,Schuurman2020,alvarez2020breakdown,Python2021}, assuming that media coverage reflects the overall public reaction. However, the topics discussed in the media are not necessarily a complete \textit{r\'epertoire} of issues that people will become interested in \cite{giani2021wait,yasseri2016}. Other quantitative work utilized election results as a proxy measurement, routinely finding a shift to the political right \cite{Berrebi2006,Bali2007,Berrebi2008,Gould2010,Abrahms2012}. While terrorism surely affects the voting decisions, it is hard to separate its real effects from the background political context. Moreover, most such studies are naturally limited to specific countries and specific times. Effects of terrorism on economics were examined, often focusing on macro level \cite{Frey2007,Benmelech2010,Becker2011,Benmelech2012}. Medium and long term shaping of public opinion after terrorist events was also studied \cite{Huff2017,bove2022effects}, but investigation of immediate and direct effects is still missing. \\[0.2cm]
To fill this gap, one would seek to identify what topics are searched online more frequently after a terrorist attack. However, the diversity of the search terms concerning any subject makes it difficult to use the search engine statistics for gauging the public interest without having a predefined list (r\'epertoire) of keywords. Instead, we employ the viewing statistics of the Wikipedia content. Researchers already shown the correlation between Wikipedia view statistics and Google search volume \cite{Ratkiewicz2010,Yoshida2015}, making it a useful proxy measure of the Web public attention \cite{garcia2016}. Wikipedia viewing statistics has in fact become an established instrument, enabling not only to comprehend but also to predict a broad variety of social phenomena  \cite{mestyan2013,yasseri2016wikipedia,yasseri2014can,ferron2011studying,moat2013,mestyan2013,Kanhabua2014,mciver2014,thalhammer2016,ban2017,Garcia2017,Candia2019,lorenz2019accelerating,ribeiro2021sudden,kobayashi2021,Arnold2021}. It can safely be argued that Wikipedia encompasses ``all'' issues of potential interest, including those ignited by terrorist attacks. As its information is organized into articles, it is convenient for mapping the public responses into certain topics and categories. Articles represent the missing r\'epertoire.\\[0.2cm]
We studied the viewing statistics of all articles in English Wikipedia around the time of 15 terrorist attacks in Europe (see section \ref{Sample} and Appendix \ref{appendix:a}). By detecting the spikes in \textit{daily viewcount} of specific articles we identified the issues that the public interest is directed towards as a consequence of these events (see section \ref{Preliminaries}). Unfortunately, this data is available only from 2015 on. This makes the Paris attack of 2015 the most suitable event to commence our analysis.


\section{Results}

\subsection{Changes of collective attention around Paris 2015 attack}

\begin{figure}[!hbt]
\centering
\includegraphics[width=\textwidth]{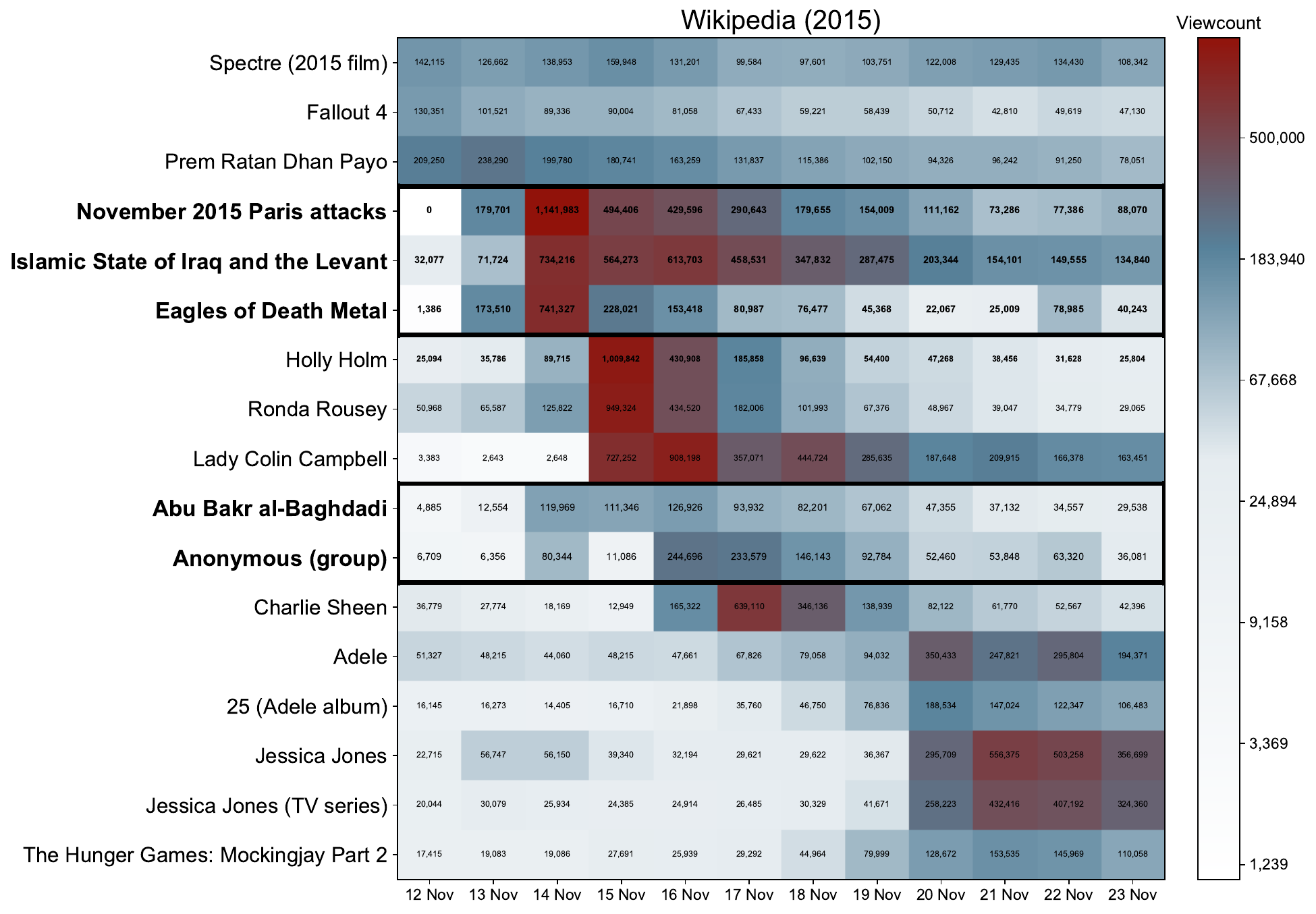}
\caption{Articles identified with excess attention around the date of the Paris attack in 2015. The color encodes the total views of a given article on a given day (also written within each cell). Articles directly related to the attack are shown in boldface font.}
\label{fig-zoom-to-paris}
\end{figure}

We commence by a thorough examination of the Paris attack in 2015, which is the initial event within our sample. We will broaden our analysis to encompass all attacks later. Focusing on the 12 days surrounding the Paris event, we identify 17 articles that received the most attention in this period, shown in Fig.~\ref{fig-zoom-to-paris} (for details on the identification method see section~\ref{SampleParis}). Among these 17 articles, we found that five were directly related to the attack while the remaining 12 were not. On the 12 November 2015, a day before the attack, the largest attention was devoted to three articles related to entertainment: Prem Ratan Dhan Payo (Hindi romantic movie released on 12 November), Spectre (2015 James Bond movie released on 26 October), and Fallout 4 (computer game released on 10 November). The attack starts on the evening of 13 November in Paris local time (Since this was still early afternoon in North America, many English-speaking views registered on 13 November, even though the event lasted until the early hours into 14 November.). \\[0.2cm]
This immediately diverts attention to the newly created article \textit{November 2015 Paris attacks} in addition to \textit{The Islamic State of Iraq and the Levant} (ISIL, the main perpetrator) and \textit{Eagles of Death Metal} (rock band that was playing in Bataclan theatre). The attention to these three articles spiked on 14 November, but without dramatically reducing the attention on the above, mentioned entertainment articles. On 15 November the collective attention splits among these three attack-related articles, the final fight of the women's Ultimate Fighting Championship (held on 14 November), and the British TV personality Lady Colin Campbell due to her appearance in a UK reality show on the same day. \\[0.2cm]
The situation is largely unchanged on 16 November, except that attention peaks for other two attack-related articles: \textit{Abu Bakr al-Baghdadi} (leader of ISIL) and the group \textit{Anonymous}, who following the attack announced a major hacking operation against ISIL. Meanwhile, the attention to three main attack-related articles declined and is overshadowed on 17 November by the news that the actor Charlie Sheen was HIV-positive. Eventually, on 20 November, attention is taken by three entertainment releases: the album ``25'' by the popular singer Adele, the detective TV series Jessica Jones, and the new part of the dystopian movie Hunger games. \\[0.2cm]
Of the five attack-related articles, by 23 November most attention is devoted to ISIL. The viewing dynamics of all articles exhibit an intense spike immediately after the triggering event followed by a gradual decay in days to follow, consistently with \cite{Candia2019,lorenz2019accelerating}. This pattern seems not to differ between attack-related and other articles.

\subsection{Collective memory of terrorist attacks}

\begin{figure}[!hbt]
\centering
\includegraphics[width=\textwidth]{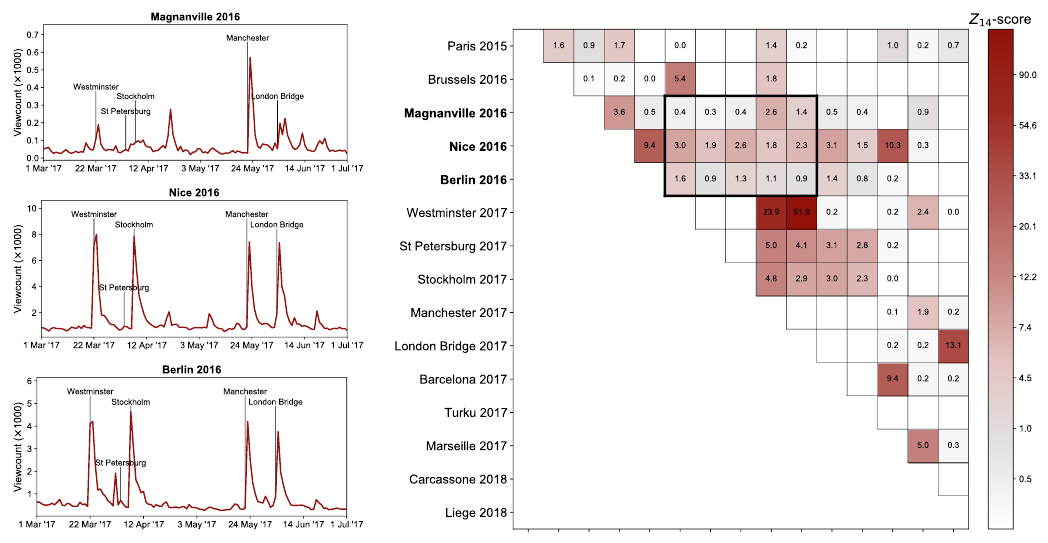}
\caption{Collective memory of terrorist attacks. Three left panels: Time series of viewcounts to articles for Magnanville 2016 (top), Nice 2016 (middle), and Berlin 2016 (bottom), shown for the period of four months from March to June 2017. The dates of five terrorist attacks that happened during that time interval are indicated. Right panel: the excess attention $Z$ was received by each attack as triggered by later attacks. Attacks are arranged vertically (and horizontally) in chronological order from top to bottom (left to right). Values of $Z$ for the three panels on the left panel are marked by a bold square.}
\label{fig-collective-memory}
\end{figure}

We now proceed to assess all 15 attacks in our sample, rather than exclusively focusing on Paris 2015. Research has shown that current events not only attract collective attention to themselves but also trigger collective memories of similar past events \cite{Garcia2017,Candia2019,lorenz2019accelerating,ribeiro2021sudden,kobayashi2021,Arnold2021}. To examine this for the case of terrorist attacks, we look at the Wikipedia articles devoted to all attacks in our sample and ask how strongly each new attack triggers the attention to the previous attacks. In the top left panel of Fig.~\ref{fig-collective-memory} we show the viewcount time series for Magnanville 2016 from March 2017 through June 2017. Five terrorist attacks that happened during that interval are indicated. Manchester 2017 made the public intensely recall the events of Magnanville 2016. London Bridge 2017 and Westminster 2017 did so somewhat less intensely, while St. Petersburg 2017 and Stockholm 2017 did not at all. Considering this along with the results from the other two left panels, we note that when a terrorist attack happens, the public is reminded of prior (similar) attacks, albeit of some more and some less. \\[0.2cm]
To study this collective memory more systematically, we focus on a quantity we call ``excess attention'' denoted with $Z$. It expresses how strongly a newer attack triggers the attention to an older attack (see section \ref{ExcessAttention}). The calculated values of the excess attention are shown in the right panel of Fig.~\ref{fig-collective-memory}. Most events have triggered considerable excess of attention to the past events, confirming the existence of a collective memory of terrorist attacks in Europe. The strongest reactions are found for pairs of attacks that happened in the same country in line with the findings by García-Gavilanes et al. \cite{Garcia2017}. Also, the time elapsed since the attack seems to be a strong factor, as attacks that happened a long time ago are seldom remembered.


\subsection{General patterns in public reaction across attacks} 

Reaction to a terrorist attack is shaped by many factors, primarily by the media portrayal of the event, but also by the perceptions established in the target society about itself and the perpetrators. Searching Wikipedia is a matter of individual choice, but this choice is founded on an array of social narratives that entail symbols, stories and memories of a given group, and turns them into a shared conceptual system \cite{Kerby1991,Nelson2003}. What narratives underpin the searching choices in Wikipedia after a terrorist act? To answer this, we identify the universal effects that attacks have on the collective attention. \\[0.2cm]
Starting from \textit{all} Wikipedia articles (section \ref{Preliminaries}), we deployed an exclusion pipeline combining time series analysis and qualitative approaches (sections \ref{ExcessAttention} and \ref{Exclusion}), to extract 69 articles to which attention is consistently drawn after terrorist attacks. We then manually grouped them into 19 \textit{specific categories} according to the commonalities of their thematic codes (section \ref{Specific}). Finally, by finding coherent patterns among specific categories we grouped them into four \textit{broad categories} (section \ref{Broad}). This arrangement is illustrated in Fig.~\ref{fig-hierarchy} and schematized in the Appendix \ref{appendix:b}. It can be considered in a hierarchy, whereby individual articles belong to one of the specific categories, and each specific category belongs to one broad category. Broad categories represent a coarse, yet robust and interpretable level of description. They are the four core overarching groups of interests that always arise after terrorist attacks. 
\begin{description}
\item[ENEMY] It is easy to recognize that the concept of \textit{enemy}, identified as a threat, underpins specific categories related to both terrorism and conspiracies. The basic construction of the enemy goes in two ways: the Islamic extremist, a person driven by extremist ideology posing a danger and/or the terrorist, an individual who is relying on terror as the method of achieving goals in the West and elsewhere. We interpret this as a natural reaction of the English-speaking public to a terrorist attack: Establish the perpetrators (Islamic extremists) as the enemy, and then learn about their methods and actions, especially in the West.
\item[SECURITY] National security, weapons, transport, insurance, and survival equipment have the idea of security as the common denominator. Seeking shelter is a natural response to the threat of the enemy. Being under attack puts us into \textit{fight or flight} mode: either to confront the enemy or escape from the threat. Fight mode corresponds to the interest in armed forces, military equipment, police, conscription, and usage/purchasing of weapons. The public wants to rely on the official national security mechanisms, but also be ready for individual defence (``take matters into their own hands''). Flight mode is found in the interest of means of transportation (ways to escape), insurance (financial and otherwise), and equipment for survival in the wild (the ultimate shelter in disruptive circumstances).
\item[OTHER-PERCEPTION] Once established who is the enemy and how to defend/escape from it, one wants to learn \textit{about} the enemy. We recognize this as the common motif in pronounced interest for Islam and Muslims, which West often perceives \cite{Said1978} as the archetype of the ``Other''. Of course, this role has deep historical and cultural roots, which quickly spring out upon a terrorist attack. Induced by it is also a curiosity about the practices of Islam, its origins and history, sacred texts, religious figures, presence in the West and elsewhere, and even tourist attractions in the Middle East. 
\item[SELF-PERCEPTION] Finally, one wants to soul-search and re-evaluate who ``We'' are. This establishes a self-perception that contrasts the perception of ``The Other'' and relies on both historical and imagined features of the West. It is centered around Western national/religious identities and political activism but also includes anti-Islam groups and figures. Interestingly, there is a large curiosity about violent historic events orchestrated by the Western powers. Given the context of the terrorist attack, it is not surprising that self-perception is built upon (equally) violent aspects of the West. On the other hand, this can be interpreted as a form of self-criticism, a realization that the West has always been violent regardless of Islamic terrorism (``they are bad, but we are no angels either''). 
\end{description}
\begin{figure}[!hbt]
\centering
\includegraphics[width=\textwidth]{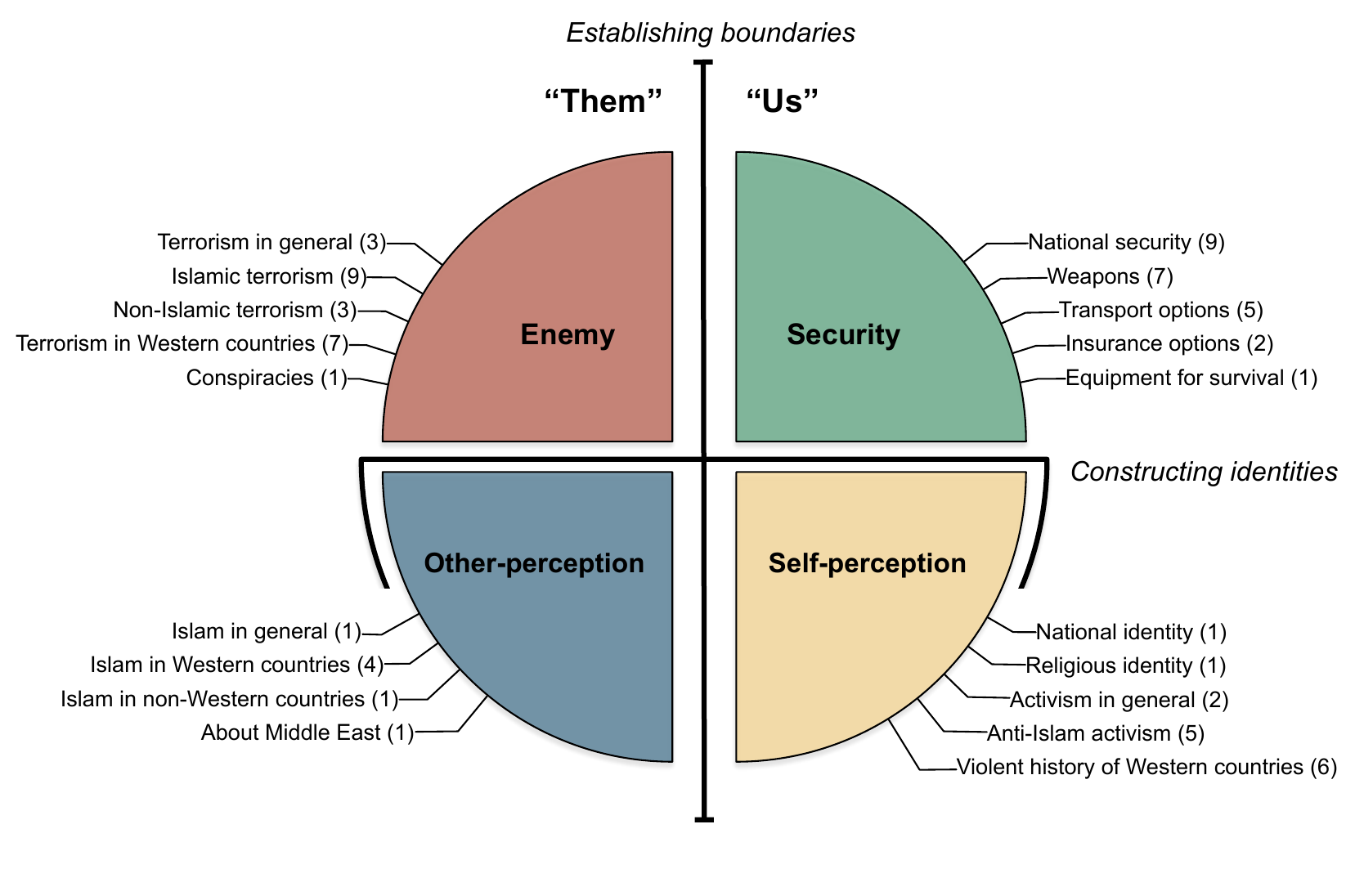}
\caption{Organization of 69 articles that survived all the filters (Table~\ref{tab-all-reacted}) into specific and broad categories (Appendix \ref{appendix:b}). Four broad categories are illustrated by different colors and positioned as to capture the relationships among them, i.e., the distinctions along two independent dimensions: Establishing boundaries and Constructing identities. Two broad categories on the left explain ``Them'', while the two on the right explain ``Us''. Specific categories are shown schematically by their names, whereby in parentheses we report the number of articles belonging to each specific category.}
\label{fig-hierarchy}
\end{figure}
The organization of the broad categories can be viewed along two independent dimensions. First, it is easy to recognize that two broad categories on the left, Enemy and Other-perception, describe ``Them'' in very general terms. Readers are compelled to learn about them, not just as the enemy, but to clarify who they are and how shall they perceive them. Readers look for a wider context of where the enemy comes from and why. In contrast, two broad categories on the right describe ``Us''. They amount to the construction of a sense of belonging and a safe community, ``Our'' community. This leads to the introspection of who they are in the face of the enemy. Both ``Us'' and ``Them'' can be recognized sharper when placed in opposition to each other. This binary distinction can be interpreted as \textit{Establishing boundaries}: Enemy vs Security, threat/danger vs safety/belonging, unknown vs known, Other vs Self, ``Them'' vs ``Us''. \\[0.2cm]
Second, two broad categories in the bottom amount to \textit{Constructing identities}. The disruptive situation makes the public interested not only in the immediate concerns but in a more comprehensive and abstract perspective on the event. The public becomes curious about Islam not exclusively from a retaliatory perspective. On the other hand, that same public is looking for historic and current facts about Western nations and their political/social organization, neither of which is immediately related to the attacks. Both arising interests point to the need for an identity, i.e., the understanding of how ``We'' perceive ourselves in face of the ``Them''. This elucidates how they establish the boundary between what they see as shared within their community in contrast to whatever is outside of it (and potentially threatening). Interestingly, public also gets curious about Islam's \textit{presence in the West} -- the place where the ``They'' meet ``Us''. \\[0.2cm]
Next, we examine the 15 terrorist attacks through lenses of excess attention $Z$. For each attack, we calculate its total excess attention by summing over all broad categories. This leads to a single value of $Z$ for each attack, reported in the left panel of Fig.~\ref{fig-Zscores-broad}. We find the most intense reaction to Paris 2015, Manchester 2017, London Bridge 2017, and Marseille 2017. Other attacks trigger weaker but still noticeable reactions. We could not find meaningful correlations between these values and quantifiable parameters surrounding the attacks, such as the number of deaths or injuries, as it was done by García-Gavilanes et al. \cite{garcia2016}. The same was found true for the size of the Muslim population in the target countries. Still, we noticed that attacks that took place in the UK and France seem to generate somewhat more excess attention than those occurring elsewhere. A major attack in Brussels in 2016 triggered comparatively weaker interest. \\[0.2cm]
Looking at the excess attention of four broad categories concerning the attacks (right panel of Fig.~\ref{fig-Zscores-broad}), we find that the attention was most intensely diverted to the Enemy category after Paris 2015, and somewhat less so after Manchester 2017 and Marseille 2017. Turku 2017 triggered no interest in the Enemy, while Barcelona 2017 and Liege 2018 exhibit very weak interest in the Enemy. We found the weakest overall reaction in the Security category, where seven attacks did not react at all. In contrast, London Bridge 2017 has an extreme reaction in this category. Redirection to the Other-perception category is somewhat related to that of the Enemy category. Many attacks show similar values for both, with Manchester 2017 and Marseille 2017 being the highest. For the Self-perception category we find moderate values across all attacks, but not much correlated with the Security. The intensities of reactions across four broad categories are unique for each event, the strongest being for Paris 2015. \\[0.2cm]
This paints the picture of each attack having a context of its own and impacting the public in a particular way. The public reaction to attacks is diverse and heterogeneous, with different attacks eliciting varying degrees of response within different broad categories. Nevertheless, we emphasize that despite these nuances, all reactions underscore a division between ``Us'' and ``Them''. Furthermore, we stress that the extent of this dichotomy varies across attacks, although we were unable to identify any external variables that could explain these variances.

\begin{figure}[!hbt]
\centering
\includegraphics[width=\textwidth]{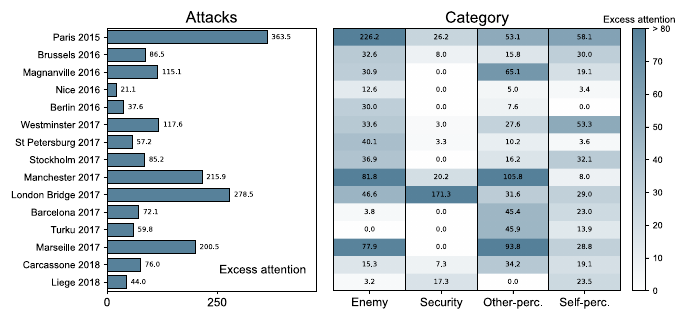}
\caption{Intensity of the public reaction due to terrorist attacks, which are shown vertically by their abbreviations (section \ref{Sample} and Appendix \ref{appendix:a}). Left panel: total Excess attention for every attack aggregated cumulatively over all of the broad categories. Right panel: total Excess attention for every attack broken down over four broad categories, which are shown horizontally. The color bar is truncated for Excess Attention $> 80$ for better clarity.}
\label{fig-Zscores-broad}
\end{figure}


\subsection{Returning the collective attention to `normal'}

\begin{figure}[!hbt]
\centering
\includegraphics[width=\textwidth]{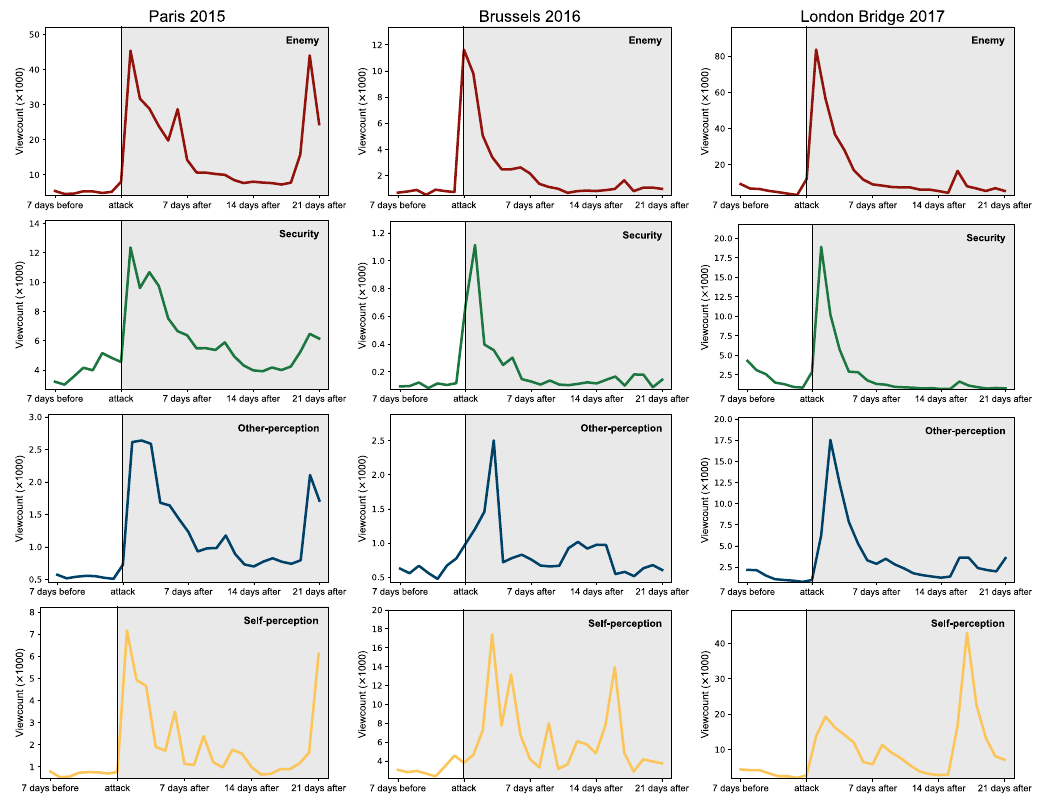}
\caption{Dynamics of decay of collective public attention after terrorist attacks. Three top row panels: the cumulative time series of viewcounts for Paris 2015, Brussels 2016 and London Bridge 2017, for the broad category Enemy. The time of each attack is marked by a vertical line and the period after it is shaded in grey. Second, third, and fourth rows of panels: the same for the other three broad categories.}
\label{fig-restoration}
\end{figure}

After an attack, how gradually is the public attention restored to routine concerns? Does this depend on the attack and/or the broad category? We look at three examples: Paris 2015, Brussels 2016, and London Bridge 2017. For each of them, we construct a time series that shows the decay of public attention to each broad category. Results for the period starting a week before the attack to three weeks after the attack are shown in Fig.~\ref{fig-restoration}.\\[0.2cm]
After every attack, the decrease pattern of collective attention is similar for broad categories of Enemy, Security, and Other-perception, but it is unique to each event. Specifically, for Paris 2015 we find the slowest decrease followed by a sudden increase three weeks after the event in all three broad categories. In contrast, the patterns for Brussels 2016 and London Bridge 2017 are very similar: the decrease is relatively constant and quick for Enemy and Security, and somewhat more gradual for Other-perception. The values three weeks after these two attacks are similar to those before the attacks. This is not the case for Paris 2015, which again confirms that it was a unique event, with effects persisting long after the event and even peaking again after three weeks. In opposition to these three broad categories, we find a different situation for Self-perception. The behavior for Paris 2015 is close to that of its other three broad categories. For Brussels 2016 and London Bridge 2017 we find a rather irregular pattern, characterized by a mix of decay and sudden spikes. These observations suggest that this broad category, pointing to introspection and soul-searching, is the longest-lasting effect of the terrorist trauma.


\section{Discussion and Conclusion}

Our findings are three-fold. First, the exhaustiveness of Wikipedia enabled us to systematically study how the public prioritizes their interests in the wake of a terrorist attack and how this arrangement of interests differs from the usual trends. Second, we precisely measured which among them receive more and which had less attention. Third, we found consistent patterns in these awoken interests, grounded in the existing ecosystem of beliefs in the affected European societies. \\[0.2cm]
The reaction of the international English-speaking public to the terrorist attacks reveals the deeply rooted binary distinction between East and West, Islam and Europe, ``Them'' and ``Us''. Amid shock, a natural human (individual and collective) reaction is to search for a \textit{community} to belong to. This community has to have a meaning and identity. But the identity is always relational and constructed in opposition to ``Others'' \cite{Jenkins1996,cohen2013symbolic}, perceived as outsiders or even enemies. The enemy serves as a \textit{mirror} offering a (real or imaginary) distinction between ``our''  group and their group. This makes ``Others'' an indispensable ingredient for creating feelings of belonging. When this becomes a part of the collective social imaginary, it further fortifies the boundaries between the groups. The construction of social identity presupposes that the in-group differs significantly from the out-group. It also requires an ongoing process of maintaining and legitimizing these boundaries. In this case, ``Us'' is a transnational society in the making, based on the characteristics, values, heritage, and history of the Anglo-American, European, Western world. The ``Western'' world has been identified in opposition to Islam  through large parts of its history -- ranging from the conflicts throughout the Middle Ages, such as the Crusades and the Ottoman expansion, to the more contemporary tensions, such as the support for Israel and the Iraq wars~\cite{Said1978}. Terrorist acts -- especially if perpetrated by the religious extremists -- are ideal mechanisms for recreating this effect. \\[0.2cm]
In conjunction with a community and knowing what is and what is not part of it, we also need an understanding of ``Us'', a narrative about \textit{who we are}. This narrative is to contrast the narrative about ``Them'', which is to explain \textit{who they are}. Both narratives are usually articulated as a series of attributes about either side, as clearly revealed by the found broad categories. Specifically, the narrative of threatening Enemy to which ``we'' respond by Security contributes to the symbolic substance of ``Us'', who are residing on ``our'' side of the boundary \cite{cohen2013symbolic}. The social imaginary of the community pertains to a discursive manner of self-understanding, emerging via stories that people tell about themselves and thus construct their collective belonging \cite{potter1996representing}. In that regard, identities serve as narratives that enable the continuity of a community, form the foundation of the social collective memory, and unify a group through space and time \cite{Halbwachs1992,Lowenthal1995}.\\[0.2cm]
The imminent threat may generate other ways of defining the enemy. A prominent human feature is to search for patterns in seemingly chaotic situations while maintaining the feeling of connection. We found this in the interest for \textit{conspiracies}, which replace a visible enemy with a manufactured one. Worrisome adherence to online conspiracy theories is a large topic within computational social science \cite{DelVicario2016,van2018connecting,friedman2021humans,hofman2021Integration}. \\[0.2cm]
The last decades saw individuals' lives grow apart from large bureaucratic institutions \cite{Berger1995}. Individualism and consumerism caused the erosion of the collective loyalties, boosting the scepticism towards the state authority. This trend has been already observed in sociological classics \cite{Durkheim1951} and accelerated by modern technology and globalization \cite{castells1996information,Beck2002,Urry2003}. Interestingly, terrorism seems to push in the opposite direction, towards re-establishing the bond with (presumably) protective institutions, operating at \textit{national} level and provided by state authority. \\[0.2cm]
We also found that intense focus on topics arising from terrorist attacks dissipates after a few weeks. Typically, it is overwhelmed by the never-yielding interest in entertainment. This is hardly surprising, given that the public is known to shorten its attention span \cite{Candia2019,lorenz2019accelerating}, even during the COVID crisis \cite{ribeiro2021sudden}. This is in agreement with one of our results, namely, that the public tends to collectively remember only the most recent attacks (cf. Fig~\ref{fig-collective-memory}). In Appendix \ref{appendix:c} we show a side result: The amount of collective attention redirected from other routine interests due to terrorism is insignificant from the perspective of overall viewing behavior on Wikipedia (about 200 to 300 million views per day with weekly and seasonal oscillations). In other words, none of the attacks was `that interesting' to divert major and lasting attention from entertainment. \\[0.2cm] 
Individuals' prioritization of topics expresses narratives that they were brought up with. These involve social memories as a basis of collective (for instance national) identity, which includes cultural norms, notions of authenticity, power relations, etc. Our findings indicate that for international English-speaking audiences these narratives are centered around security, enemy, self-perception, and other-perception. In them we recognize a \textit{search for belonging and meaning} in a globalizing late modern society, triggered by a sudden disruption. These results are in accord with the classical studies on the relationship between Europe (the West) and its neighbours (the East), referring especially to the collective perceptions of the ``significant other" \cite{Said1978}. They call for reconsidering the classical ideas of nationalism, thus advancing the constructivist understanding of society \cite{cohen2013symbolic}, which is detached from the actual physical places \cite{Appadurai1996}. In that regard, we are consistent with the contemporary scholarly consideration of belonging, which is increasingly shifting towards transnational and virtual social reality, articulated through imagined communities and forged via ``fragile communication bonds'' such as common usage of English Wikipedia \cite{anderson2006,Barth1998,Delanty2018}. \\[0.2cm]
Our results could benefit from an independent verification via sources other than Wikipedia, primarily involving social media. Still, the popularity of any tweet could be due to lack of a systematic r\'epertoire of tweets, rather than genuine public interest in that tweet's content. On the other hand, what else, besides terrorist attacks, diverts attention to the 69 articles from Table~\ref{tab-all-reacted}? Perhaps some articles react \textit{only} to the terrorist attacks and could serve as `markers'. We examined this in the Appendix \ref{appendix:d} and found no evidence for them. Another methodological challenge is the robustness against the modifications of exclusion criteria. We check this, and found that key results are robust against all meaningful modifications. For more on this and other robustness checks see Appendix \ref{appendix:e}. \\[0.2cm]
There are two main directions for advancing the presented work. First, spontaneous reaction to disruptive situations naturally involves languages other than English, both in Wikipedia and elsewhere. This makes online sources in other languages prime target of future work. Second, from how we made our sample of attacks, it just so turned out that all 15 of them were committed by Islamic extremists. But what about terrorist acts perpetrated, for example, by Western far-right groups? We fully acknowledge that terrorism can have very diverse origins and that public reactions to it must be investigated. In this light, it is also of interest to look at a wider time span, starting for example from Twin Towers attack in 2001. \\[0.2cm]
Our conclusion is that while unfortunately terrorism is not likely to vanish soon, studies like this one can help understand the intricacies of the public reactions to it, and minimize the consequential societal damages such as extremism and polarization.


\section{Data and methods} 

\subsection{The sample of terrorist attacks} \label{Sample}

\begin{figure}[!hbt]
\centering
\includegraphics[width=0.65\textwidth]{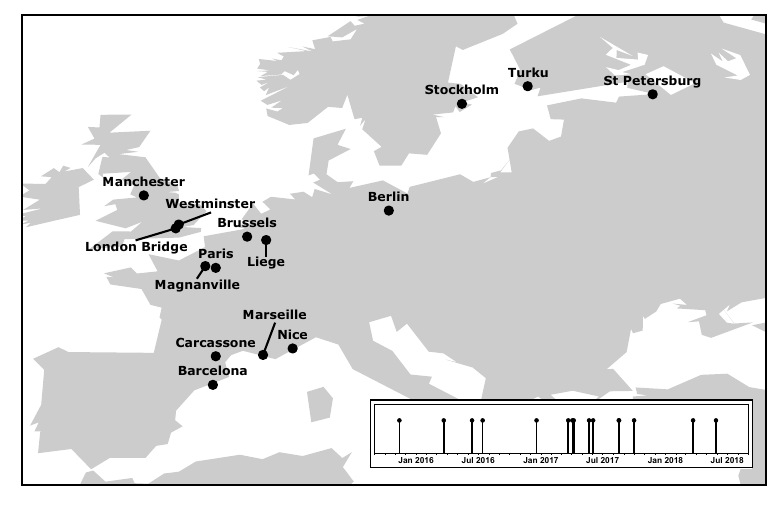}
\caption{Locations (cities) of 15 terrorist attacks comprising our sample. Attacks occurred between November 2015 (Paris) and May 2018 (Liege), as illustrated in the inset. See Appendix \ref{appendix:a} for a comprehensive overview and details of each attack.}
\label{fig-sample-of-attacks}
\end{figure}

We considered all confirmed acts of terrorism in Europe starting from November 2015 Paris attacks until the end of 2018. Using Wikipedia's information, we found 43 such acts but kept only those with at least two casualties excluding the attackers. This led us to a sample of 15 terrorist attacks across Europe (See Fig.~\ref{fig-sample-of-attacks} for the geographical location and the timeline of the incidents). Note that the available Wikipedia viewership logs commence from July 2015. This restricts our options in selecting the attacks and makes Paris 2015 the first event to start from. Of course, had we had the access to data prior to 2015, we would have started our analysis from the New York attack of 2001 and possibly broaden it beyond Europe. 

\subsection{Data preparation} \label{Preliminaries}

Our overarching aim is to examine the visits to \textit{all} articles in English Wikipedia during the relevant period and identify those articles that are visited more frequently (receive excess attention) after a terrorist attack. We started by establishing a list of entries that can be considered valid articles \cite{wikipediaNameDumps,chunksWiki}. As of August 2019, this list contained 14,733,158 entries. Through Wikimedia API \cite{wikimediaApi} we can obtain the exact viewcount (number of visits) for an article on any day. This \textit{daily viewcount} is a great proxy for the attention devoted to that topic by the international English-speaking audience. We retrieved the time series of daily viewcounts for the entire list with 14,733,158 entries from 1 January 2015 to 8 August 2019 (1680 days). Each value is the total number of visits to an article on a given day, cumulatively generated by entire online traffic from all devices with access to the Internet. We want to examine only those articles whose viewcounts spiked after terrorist attack(s) and whose content can be meaningfully connected to the attack. After the extensive exclusion procedure described below, we identified 69 such articles.


\subsection{Sampling the articles around the time of Paris 2015 attack} \label{SampleParis}

We sample the articles for Fig.~\ref{fig-zoom-to-paris} by considering the 12 days surrounding the event. We found this time window most suitable for illustrating the dynamics of collective attention around the attack. We looked at the top 5 articles by viewcount on each of these 12 days. We then selected articles, that on at least two days (of these 12 days) received more than 100,000 views. This gave us the sample of 17 articles, shown vertically in Fig.~\ref{fig-zoom-to-paris}.


\subsection{Definition of Excess attention} \label{ExcessAttention}

Excess Attention, $Z$, is defined as follows. Let $A$ be a Wikipedia article and $T$ the time of a specific attack. The mean $\mu (A,T)$ and standard deviation $\s (A,T)$ of \vs of $A$ before the attack are calculated starting from 1st of January 2015 as:
\[ \mu (A,T) = \frac{1}{\textit{T-1}}\sum_{\textit{t=1}}^{\textit{T-1}} w (A,t) \;\; , \;\;\;\;\;\;
  \s (A,T) = \sqrt{\frac{1}{\textit{T-1}}\sum_{t=1}^{T-1} \Big( w (A,t) - \mu (A,T) \Big)^2} \;\; , \]
where $w (A,t)$ is the \vi of $A$ on the day $t$. Next, we compute $\nu (A,T)$, the average \vi of $A$ in the week following the attack:
\[ \nu (A,T) = \frac{1}{7} \Big( w (A,T) + w (A,T+1) + \hdots + w (A,T+6) \Big) \;\; . \]
Our choice of a week (7 days) as the period for averaging is motivated by pronounced weekly oscillations of overall viewcounts and in line with \cite{garcia2016} (see also Appendix \ref{appendix:c}). Now we define $Z (A,T)$ as: 
\begin{equation} 
  Z (A,T) = \frac{\nu (A,T) - \mu (A,T)}{\s (A,T)} \;\; . 
  \label{eq-Z}
\end{equation}
Excess Attention measures the increase of mean \vs in a week after the event compared to the mean \vi before the attack, expressed in the units of standard deviation characterizing the trend prior (Our measure of Excess Attention is conceptually similar to $Z$-score, even if it is not exactly the same.) It can be computed for any article $A$ in association with any attack $T$. For example, if an article displays $Z=3$ in association with $T$, it means that in the week following $T$ the visits to $A$ have increased by 3 standard deviations compared to the average before $T$. This is our main tool in identifying the articles that have reacted to an attack in contrast to those that did not. \\[0.2cm]
For Fig.~\ref{fig-collective-memory} (right panel), since we need a more stable estimates of $\mu$ and $\s$, we exclude the first 2 weeks after the older incident. For Fig.~\ref{fig-Zscores-broad} (left panel), we assigned the total (cumulative) $Z$ to each attack by summing $Z$s for all articles that reacted to that attack, cumulatively across all broad categories. We did the same for Fig.~\ref{fig-Zscores-broad} (right panel), but summarizing each broad category separately.


\subsection{Article exclusion steps} \label{Exclusion}

\paragraph{Step 1: Removal of stubs and redirects.} Wikipedia uses the term ``stub'' for abnormally incomplete articles. Using the stub tags we identified and removed them. We also removed articles that only carry an alternative title and a link to the original article with the more conventional title (``redirects''). Starting from the complete list of articles in English Wikipedia, excluding all stubs and redirects left us with 5,876,878 articles.

\begin{figure*}[!hbt]
\centering
\includegraphics[width=0.6\linewidth]{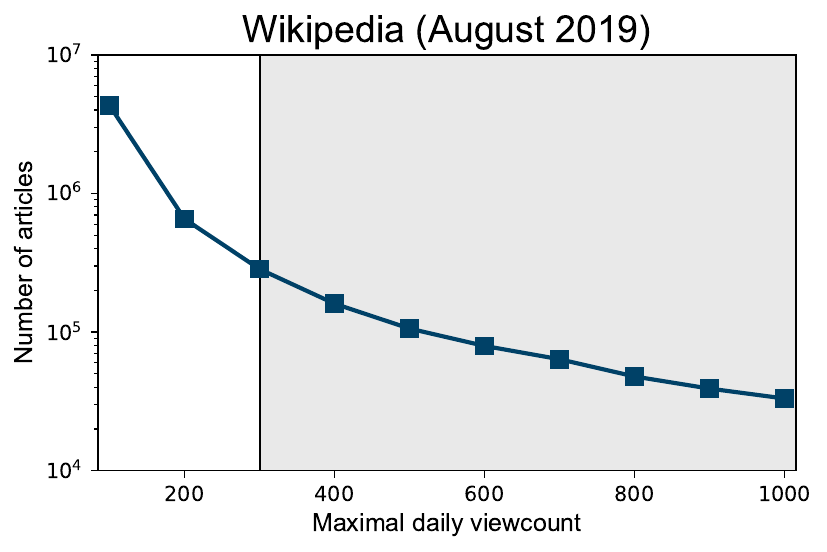}
\caption{The distribution of maximal daily viewcounts (maximal number of views reached on a single day) between 1 January 2015 and 8 August 2019, for all 5,876,878 articles after Step 1. We kept only the articles in the shaded area, i.e., excluded those that were never visited more than 300 times on a single day. Note a slight change of slope near 300, making this a natural choice for cut-off.}
\label{fig_maximal_daily}
\end{figure*}

\paragraph{Step 2: Exclusion of articles with low viewership.} To filter our low view articles, we used a cut-off based on the maximal daily viewcount. We kept only the articles whose daily viewcount at least once reached 300. See Fig.~\ref{fig_maximal_daily} for the distribution of maximum daily views and our cut-off determined based on the ``elbow rule''. This step left us with 912,192 articles.

\begin{figure}[!hbt]
\centering
\includegraphics[width=12cm]{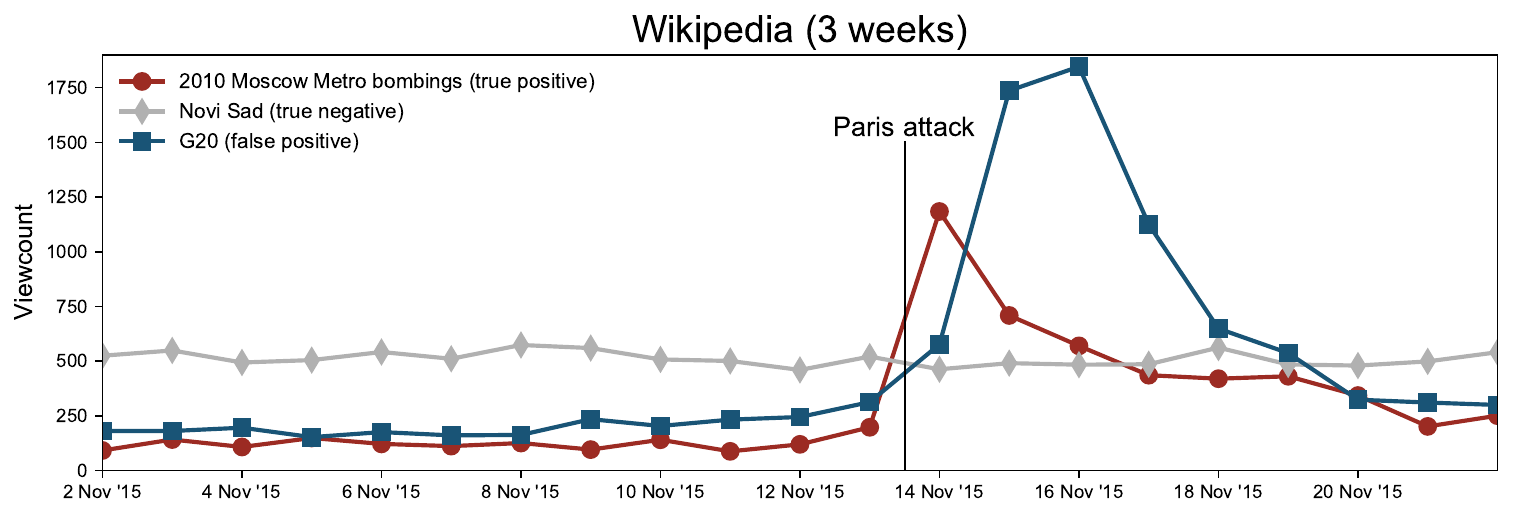} 
\caption{Viewcounts of three articles around the time of the November 2015 Paris attack. Attention on the article in red peaks immediately after the attack, and given its content, we reasonably expect that this was a reaction to the attack (true positive). The article in grey exhibits no qualitative change, and indeed, given its topic (a city in Serbia), we did not expect any (true negative). The attention is devoted to the example in blue peaks, but this article's content precludes us to attribute this to the attack (false positive).}
\label{fig-illustration-Z}
\end{figure}

\paragraph{Step 3: Removal of articles with small Excess Attention after attacks.} Next, for each attack separately, we exclude articles that received no marked increase in attention immediately after the incident (true negatives). As threshold we use $Z=3$: Assuming that \vs are normally distributed, the chance of observing $Z \ge 3$ at random is $p \sim 0.001$ (see also Appendix \ref{appendix:e}). Upon excluding articles with $Z<3$, we were left with 9,000 to 24,000 articles per attack. To better illustrate this step, we show in Fig.~\ref{fig-illustration-Z} the \vs for three example articles around the time of Paris 2015. This filter is successful in detecting articles whose increase in \vs is related to the attack (true positive), but unsuccessful in filtering out articles that had an increase in \vs due to other unrelated events (false positive). An example is G20 that received spike in attention due to G20 meeting that accidentally took place around the same time as Paris 2015. We deal more with false positives later. 

\paragraph{Step 4: Automated removal based on content} Inspecting the articles which passed the previous filters, we noticed common patterns in their topics, pointing to an overarching picture of the arising interests. Still, as too many events occur simultaneously, there were still too many false positives. To remove them, we employed semi-supervised Google's text-miner BERT (Bidirectional Encoder Representations from Transformers) \cite{devlin2018bert,bertDocumentation}. It outputs a score that expresses how similar is the content of some text to the content of the training text. We prepared the training text by taking a random sample of 10\% of articles for each attack and manually filtering out the articles which were not conceptually related to the attacks. We then extracted the summaries (Wikipedia calls `summary' a brief description of the article's content, available immediately after the title.) of all remaining articles and trained BERT on them. Then, we had BERT recognize the similar content in the summaries of all articles surviving Step 3. Upon completion, we were left with 1000 to 8000 articles per attack. We evaluated BERT's performance by manually inspecting a 10\% sample of thus excluded articles and indeed found them all true negatives.

\begin{figure}[!hbt]
\centering
\includegraphics[width=0.9\linewidth]{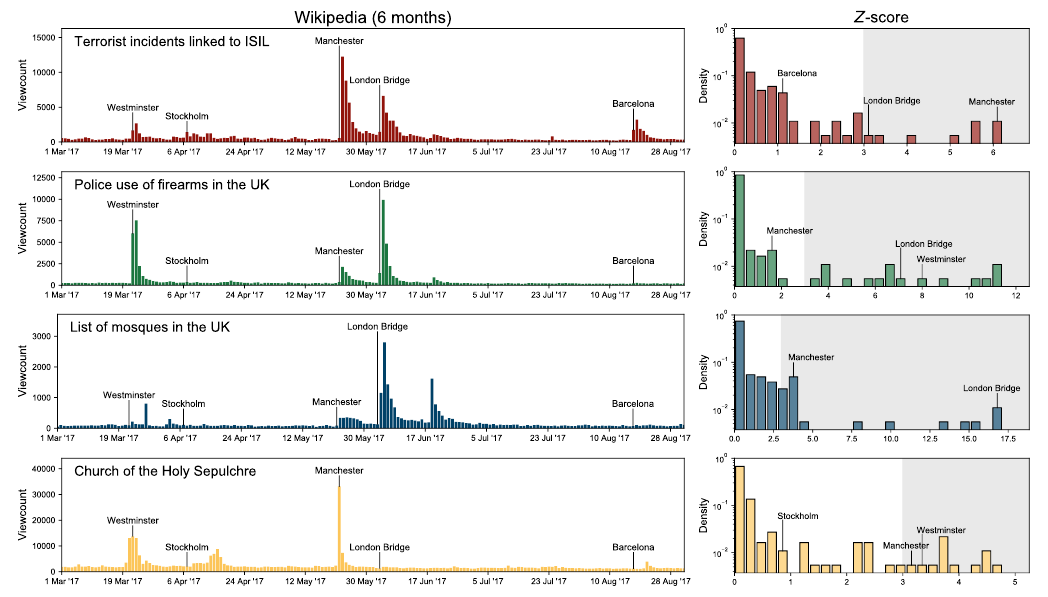}
\caption{Examples of four articles identified in Step 5. Top left: \vi time series of \textit{Terrorist incidents linked to ISIL} from 1 March 2017 to 1 September 2017. Attacks that occurred during those 6 months are indicated. Top right: the histogram of $Z$-scores for that article for the entire time span (region $Z>3$ is shaded). Other panels: same as on the top panels (left and right) for other three examples of articles. These plots demonstrate how well the threshold $Z=3$ separates day-to-day fluctuations from what we can consider a reaction to an attack.}
\label{fig-four-example-time-series}
\end{figure}

\paragraph{Step 5: Distilling the reactions.} Our next aim is to identify themes that systematically appear in public interests and are not connected to a single attack. We put the surviving articles back together and remove those that reacted to less than three independent attacks, not counting as independent the reactions to events in the same week. In establishing what constitutes a reaction to an attack, we made two additional requirements: 
\begin{enumerate}
  \item Looking at some obvious false positives, we noticed that their \vs start increasing before the attack and hence clearly \textit{not due to} the attack. To account for this we introduce $Z_0 (A,T) = Z(A,T-7)$, the $Z$-score for a week immediately before the attack. We found $Z_0$ to be unusually large for most false positives. Yet, a slightly increased $Z_0$ is not problematic as long as it is followed by a considerably higher $Z$ post-attack. For consistency with earlier steps, we requested 
  \[ Z (A,T) > 2 + \frac{3}{3-Z_0 (A,T)} \;\; \]
  for every reaction of $A$ to $T$. Hence, for an article with $Z_0 = 1$ we request at least $Z=3.5$, for articles with $Z_0 = 2$ at least $Z=5$, etc. 
  \item Since the mean $\mu (A,T)$ appears in the denominator of Eq.~\ref{eq-Z}, $Z$ is more volatile when $\mu (A,T)$ is small. An unusually high $Z$ found for $A$ with small $\mu$ could easily be an artifact (division by a number very close to zero). To account for this, we excluded the reactions of articles $A$ to attacks $T$ whenever we found $\mu (A,T) < 20$. 
\end{enumerate}
See Fig.~\ref{fig-four-example-time-series} for illustration. This step left us with 86 articles.  

\paragraph{Step 6: Qualitative (manual) exclusion based on content.} We next inspected a broader context of every remaining article around the time of each observed reaction. We assigned a thematic code to each article using open coding \cite{Flick2020} based on two criteria:
\begin{itemize}
 \item consideration of the articles’ content and its thematic relevance from the perspective of each particular attack,
 \item timing of its appearance in terms of accidental coinciding with other unrelated but simultaneously occurring events (possible false positives).
\end{itemize}
We excluded 17 articles as we found them to coincide with: sporting events; launching of movies and TV shows; music events, personalities and their media coverage; and other accidentally coinciding events (solar eclipse, legislation procedures, cultural events, etc.). There were three articles on the Syrian civil war: the Palmyra offensive in March 2016, the Russian offensive in November 2015, and the US response to the alleged chemical attack by the Syrian regime in April 2017. Upon very close inspection we could not connect them to the terrorist events in Europe. This left us with 69 articles, shown in Table~\ref{tab-all-reacted}.
\begin{table}[!hbt]
\centering
\tiny
\hspace*{-0.64cm}\begin{tabular}{| l | l | l |}
\hline
Andrew O'Hagan & Kyle Kulinski & Devin Nunes \\
\hline
James Wickstrom & Paul Weston (politician) & Sebastian Gorka \\
\hline
Tommy Robinson (activist) & False flag & Orvis \\
\hline
TD Banknorth & Thornburg Mortgage & Moinuddin Chishti \\
\hline
Islam in Qatar & Ed Husain & Islam in Poland \\
\hline
List of mosques in the United Kingdom & Quilliam (think tank) & Susa \\
\hline
Juan Carlos I & Conscription in the United States & Crisis actor \\
\hline
Kimberly Dozier & List of police firearms in the United Kingdom & Police use of firearms in the United Kingdom \\
\hline
Ring of steel (London) & Spanish Armed Forces & UK Threat Levels \\
\hline
USS Cheyenne (SSN-773) & Church of the Holy Sepulchre & Lone wolf (terrorism) \\
\hline
Mass shooting & Massacre & Bhagwan Shree Rajneesh \\
\hline
List of terrorist incidents in Australia & List of terrorist incidents in Great Britain & List of terrorist incidents in London \\
\hline
Terrorism in Europe & Terrorism in Greece & Terrorism in Sweden \\
\hline
2002 Los Angeles International Airport shooting & 2004 SuperFerry 14 bombing & 2007 Glasgow Airport attack \\
\hline
2010 Moscow Metro bombings & Abduction of Russian diplomats in Iraq & List of terrorist incidents linked to ISIL \\
\hline
Muslim Brotherhood & Rick Leventhal & Terrorism in the Philippines \\
\hline
Basque conflict & Basque National Liberation Movement & Timeline of Real Irish Republican Army actions \\
\hline
Big Sky Airlines & Boston-Maine Airways & Corporate Air \\
\hline
Dakota, Minnesota and Eastern Railroad & Eclipse Aviation & Charles B. McVay III \\
\hline
Grdelica train bombing & Greek War of Independence & Legitimacy of the NATO bombing of Yugoslavia \\
\hline
List of events named massacres & The Man with the Iron Heart & Assault weapon \\
\hline
Comparison of the AK-47 and M16 & Gun culture in the United States & Gun politics in the United States \\
\hline
Jews for the Preservation of Firearms Ownership & United States v. Miller & Universal background check \\
\hline
\end{tabular}
\caption{The list of 69 English Wikipedia articles that have reacted to at least three independent attacks. They jointly represent the content to which terrorist acts divert public attention (irrespective of the particular attack's details).}
\label{tab-all-reacted}
\end{table}


\subsection{Construction of specific categories}  \label{Specific}

Table~\ref{tab-all-reacted} is a succinct list of what the public becomes interested in following a terrorist attack. Not surprisingly, it includes terrorism, weapons, and anti-Islamism, but also national identity, the violent history of the West, and conspiracy theories. As many articles have similar content, we used axial coding \cite{Flick2020} to extract their thematic codes and connect them. By this, we manually cluster the articles into meaningful groups, defined by the coherency of their content. We call these groups \textit{specific categories} and assign a name to each of them that summarizes the content. The 69 articles from Table~\ref{tab-all-reacted} are arranged into the following 19 specific categories. 
\begin{enumerate}
\item \texttt{Terrorism in general.} Definitional aspects of terrorism, such as what constitutes a mass shooting and what is a massacre (3 articles).
\item \texttt{Islamic terrorism.} Past Islamic terrorist acts and organizations, regardless of location, context, and targets (9 articles).
\item \texttt{Non-Islamic terrorism.} Far-right and far-left terrorism and extremism in the West, not related to Islam (3 articles).
\item \texttt{Terrorism in Western countries.} Summaries of terrorist activities in the West, including those not part of our sample, regardless of religious/ideological origin (7 articles).  
\item \texttt{Conspiracies.} Possible conspiracies behind or related to the attacks (1 article).
\item \texttt{National security.} Western security and defence institutions, their official weaponry, conscription policies and related legal frameworks (9 articles).
\item \texttt{Weapons.} Weapons for personal defence and comparisons among them (e.g. AK47 vs. M16), firearms legislation and purchasing regulations (7 articles).
\item \texttt{Transport options.} Public transportation infrastructure, including air travel, railroads, and highways (5 articles).
\item \texttt{Insurance options.} Financial security and purchasing of insurance (2 articles).
\item \texttt{Equipment for survival.} Purchasing equipment for surviving in the wild (1 article).
\item \texttt{Islam in general.} Islam religious figures (1 article).
\item \texttt{Islam in Western countries.} Muslim religious institutions, figures, and practices in the West (4 articles).
\item \texttt{Islam in Non-western countries.} Role and history of Islam in non-Western countries (1 article).
\item \texttt{About Middle East.} Locations and history of the Middle East (1 article).
\item \texttt{National identity.} Generalities about Western nations and their political order (1 article).
\item \texttt{Religious identity.} Christian religious institutions in the West (1 article).
\item \texttt{Activism in general.} Western political activists and writers, not related to Islam (2 articles).
\item \texttt{Anti-Islam activism.} Western far-right politicians, public figures and anti-Islam extremists (5 articles).
\item \texttt{Violent history of Western countries} Violent historic events in the West, famous battles in the two World Wars, local wars of independence (6 articles).
\end{enumerate}
Arrangement of individual articles from Table~\ref{tab-all-reacted} into above 19 specific categories is shown in form of table in Appendix \ref{appendix:b}.


\subsection{Construction of broad categories} \label{Broad}

The above organization of articles into specific categories is not immune to ambiguities, such as placing an article into one category or another or merging or splitting categories. We resolve this by noticing similarities across specific categories' thematic codes. For instance, categories 1--4 revolve around terrorism, whereas categories 11--14 deal with varying aspects of Islam and its relation to (real or perceived) perpetrators. This allows us to coarse-grain 19 specific categories into four \textit{broad categories}: Enemy (Specific categories 1--5, 23 articles), Security (Specific categories 6--10, 24 articles), Other-perception (Specific categories 11--14, 7 articles), and Self-perception (Specific categories 15--19, 15 articles). \\[0.2cm]
We verified that the topical coherency of broad categories is robust against the above ambiguities and even against the most meaningful modifications of the exclusion choices (e.g. requesting the reaction to only two independent attacks). To further increase the reliability and validity, we sought to define their thematic codes in the context of discourse analysis. We applied intra-coding in terms of re-checking broad categories' thematic codes 12 months after the initial coding process and found them to be consistent (We decided against deploying inter-coded reliability since we did not find it compatible with recursive and incremental nature of our coding process.). Thus constructed broad categories are \textit{universal} and interpretable as the broad public interests awakened by any terrorist act.


\section*{Acknowledgements}

We acknowledge the support of the Ministry of Education, Science and Technological Development of the Republic of Serbia (Grant No. 451-03-68/2022-14/200125) and by the Slovenian Research Agency (ARRS) via programs P1-0383 and P5-0168, and via project J5-8236. In memory of our dear colleague Novica Bjelica, who helped us set up the datasets.

\section*{Data availability}

The datasets behind this study are freely available from Wikipedia. For easier access by the scientific community, we deposited them at ZENODO repository, at \url{https://zenodo.org/record/7198794#.Y0mCXUpBxkg}. We offer our consultation to whoever wishes to work on this data.

\newpage
\begin{appendices}

\section{Details of our sample of terrorist attacks} \label{appendix:a}

\begin{table}[!hbt]
\begin{center}
\resizebox{\textwidth}{!}{ 
\begin{tabular}{ | p{2cm} | p{5cm} | p{9cm} | p{1.5cm} | p{1.5cm} | }
\hline
	Date & Wiki article (abbreviation) & Details &  Deaths & Injuries \\ \hline
	13-Nov-15 & November 2015 Paris attacks\cite{2015ParisAttack} \newline (Paris 2015) & A series of co-ordinated attacks. The first shooting attack occurred in a restaurant and a bar in the 10th arrondissement of Paris. Other bombings took place outside the Stade de France stadium in the suburb of Saint-Denis. 
 & 130 (+7\newline attackers) & 413 \\ \hline
	22-Mar-16 & 2016 Brussels bombings\cite{2016BrusselsAttack}\newline (Brussels 2016) & There were three coordinated suicide bombings in Brussels: two at Brussels Airport in Zaventem, and one at Maalbeek metro station. & 32 (+3 \newline attackers) & 340 \\ \hline
	13-Jun-16 & 2016 Magnanville stabbing\cite{2016MagnanvilleStabbing} \newline (Magnanville 2016) & A man stabbed and killed a police officer in his home, before taking the officer's wife and son hostage. ISIL claimed responsibility. & 2 (+1 \newline attacker) & 0 \\ \hline
	14-Jul-16 & 2016 Nice attack\cite{2016NiceTruckAttack}\newline (Nice 2016) &  A cargo truck was deliberately driven into crowds celebrating Bastille Day on the Promenade des Anglais in Nice. ISIL claimed the responsibility. & 86 (+1 \newline attacker) & 458 \\ \hline
	19-Dec-16 & 2016 Berlin attack\cite{2016BerlinTruckAttack}\newline (Berlin 2016) & A truck was driven into a Christmas market in Berlin. ISIL claimed responsibility. & 12 & 56 \\ \hline
	22-Mar-17 & 2017 Westminster attack\cite{2017WestminsterAttack} \newline
	(Westminster 2017) & A Muslim convert drove a car into pedestrians on Westminster Bridge, He then crashed his car into the fence of the Palace of Westminster and fatally stabbed an unarmed policeman. & 5 (+1 \newline attacker) & 50 \\ \hline
	03-Apr-17 & 2017 Saint Petersburg Metro bombing\cite{2017SaintPetersburgMetroBombing}\newline (St Petersburg 2017) & A suicide bomber blew himself up on the St Petersburg Metro, Imam Shamil Battalion, an Al-Qaeda affiliate, claimed responsibility, but according to the FSB, attacker acted on the orders of a field commander from ISIL. & 5 (+1 \newline attacker) & 64 \\ \hline
	07-Apr-17 & 2017 Stockholm attack\cite{2017StockholmTruckAttack}\newline (Stockholm 2017) & A hijacked truck was driven into pedestrians along a shopping street before crashing into a department store. The attacker had shown sympathies for extremist organizations including ISIL. & 5 & 14 \\ \hline
	22-May-17 & Manchester Arena bombing\cite{2017Manchester}\newline (Manchester 2017) & A suicide bombing was carried out at Manchester Arena after a concert by American singer Ariana Grande. & 22 (+1 \newline attacker) & 512 \\ \hline
	03-Jun-17 &  2017 London Bridge attack\cite{2017LondonBridgeAttack}\newline (London Bridge 2017) & A van ran into pedestrians on London Bridge and then drove to Borough Market. & 8 (+3 \newline attackers) & 48 \\ \hline
	16-Aug-17 & 2017 Barcelona attacks\cite{2017BarcelonaAttacks}\newline (Barcelona 2017) & Two suspects were killed in an initial accidental explosion during the preparation of explosives. Latter on, a van was driven into pedestrians in Las Ramblas, Barcelona. The following day a related attack occurred in Cambrils. ISIL claimed responsibility. & 16 (+8 \newline attackers) & 152 \\ \hline
	18-Aug-17 & 2017 Turku stabbing\cite{2017TurkuAttack}\newline (Turku 2017) & An ISIL inspired attacker said a motive for his attack was airstrikes by the Western Coalition during the 2017 Battle of Raqqa in Syria.  & 2 & 8 (+1 \newline attacker) \\ \hline
	01-Oct-17 & Marseille stabbing\cite{2017MarseilleStabbing} \newline (Marseille 2017) & Two women were stabbed by a migrant from Tunisia. ISIL claimed responsibility.  & 2 (+1  \newline attacker) & 0 \\ \hline
	23-Mar-18 & Carcassonne and Trèbes attack\cite{2018Carcassonne}\newline (Carcassonne 2018) & An attacker stole a car, killing a passenger in Carcassonne. Then, he attacked a supermarket. & 4 (+1 \newline attacker) & 15 \\ \hline
	29-May-18 & 2018 Liège attack\cite{2018LiegeAttack} \newline (Liege 2018) & A man killed two police officers and a civilian. He is also believed to have killed a man the day before the attack. & 4 (+1 \newline attacker) & 4 \\ \hline
\end{tabular}}
\end{center}
\caption*{Overview and descriptions of 15 terrorist attacks comprising our sample and the abbreviations used in the main text. Dates are as established for our analysis, although some attacks spanned several days.}
\end{table}

\newpage

\section{Arrangement of surviving 69 articles into Specific and Broad categories} \label{appendix:b}

\begin{table}[!hbt]
\centering
\tiny
\hspace*{-0.64cm}
\begin{tabular}{|c|c|l|}
\hline
\textbf{Broad category}         & \textbf{Specific category}   & \textbf{Article from English Wikipedia} \\
\hline
\multirow{23}{*}{\textbf{ENEMY}}                                                     & \multirow{3}{*}{Terrorism in general}                                                            & Massacre                                        \\ \cline{3-3} 
                                                                                     &                                                                                                  & Mass shooting                                   \\ \cline{3-3} 
                                                                                     &                                                                                                  & Lone wolf (terrorism)                           \\ \cline{2-3} 
                                                                                     & \multirow{9}{*}{Islamic terrorism}                                                               & 2002 Los Angeles International Airport shooting \\ \cline{3-3} 
                                                                                     &                                                                                                  & 2004 SuperFerry 14 bombing                      \\ \cline{3-3} 
                                                                                     &                                                                                                  & 2007 Glasgow Airport attack                     \\ \cline{3-3} 
                                                                                     &                                                                                                  & 2010 Moscow Metro bombings                      \\ \cline{3-3} 
                                                                                     &                                                                                                  & Abduction of Russian diplomats in Iraq          \\ \cline{3-3} 
                                                                                     &                                                                                                  & List of terrorist incidents linked to ISIL      \\ \cline{3-3} 
                                                                                     &                                                                                                  & Muslim Brotherhood                              \\ \cline{3-3} 
                                                                                     &                                                                                                  & Rick Leventhal                                  \\ \cline{3-3} 
                                                                                     &                                                                                                  & Kimberly Dozier                                 \\ \cline{2-3} 
                                                                                     & \multirow{3}{*}{Non-Islamic terrorism}                                                           & Basque conflict                                 \\ \cline{3-3} 
                                                                                     &                                                                                                  & Basque National Liberation Movement             \\ \cline{3-3} 
                                                                                     &                                                                                                  & Timeline of Real Irish Republican Army actions  \\ \cline{2-3} 
                                                                                     & \multirow{7}{*}{Terrorism in Western countries}                                                  & List of terrorist incidents in Australia        \\ \cline{3-3} 
                                                                                     &                                                                                                  & List of terrorist incidents in Great Britain    \\ \cline{3-3} 
                                                                                     &                                                                                                  & List of terrorist incidents in London           \\ \cline{3-3} 
                                                                                     &                                                                                                  & Terrorism in Europe                             \\ \cline{3-3} 
                                                                                     &                                                                                                  & Terrorism in Greece                             \\ \cline{3-3} 
                                                                                     &                                                                                                  & Terrorism in Sweden                             \\ \cline{3-3} 
                                                                                     &                                                                                                  & Terrorism in the Philippines                    \\ \cline{2-3} 
                                                                                     & Conspiracies                                                                                     & False flag                                      \\ \hline
\multirow{24}{*}{\textbf{SECURITY}}                                                  & \multirow{9}{*}{National security}                                                               & Conscription in the United States               \\ \cline{3-3} 
                                                                                     &                                                                                                  & Crisis actor                                    \\ \cline{3-3} 
                                                                                     &                                                                                                  & List of police firearms in the United Kingdom   \\ \cline{3-3} 
                                                                                     &                                                                                                  & Police use of firearms in the United Kingdom    \\ \cline{3-3} 
                                                                                     &                                                                                                  & Ring of steel (London)                          \\ \cline{3-3} 
                                                                                     &                                                                                                  & Spanish Armed Forces                            \\ \cline{3-3} 
                                                                                     &                                                                                                  & UK Threat Levels                                \\ \cline{3-3} 
                                                                                     &                                                                                                  & Universal background check                      \\ \cline{3-3} 
                                                                                     &                                                                                                  & Devin Nunes                                     \\ \cline{2-3} 
                                                                                     & \multirow{7}{*}{Weapons}                                                                         & Comparison of the AK-47 and M16                 \\ \cline{3-3} 
                                                                                     &                                                                                                  & Gun culture in the United States                \\ \cline{3-3} 
                                                                                     &                                                                                                  & Gun politics in the United States               \\ \cline{3-3} 
                                                                                     &                                                                                                  & Jews for the Preservation of Firearms Ownership \\ \cline{3-3} 
                                                                                     &                                                                                                  & Assault weapon                                  \\ \cline{3-3} 
                                                                                     &                                                                                                  & USS Cheyenne (SSN-773)                          \\ \cline{3-3} 
                                                                                     &                                                                                                  & United States v. Miller                         \\ \cline{2-3} 
                                                                                     & \multirow{5}{*}{Transport options}                                                               & Big Sky Airlines                                \\ \cline{3-3} 
                                                                                     &                                                                                                  & Boston-Maine Airways                            \\ \cline{3-3} 
                                                                                     &                                                                                                  & Corporate Air                                   \\ \cline{3-3} 
                                                                                     &                                                                                                  & Dakota, Minnesota and Eastern Railroad          \\ \cline{3-3} 
                                                                                     &                                                                                                  & Eclipse Aviation                                \\ \cline{2-3} 
                                                                                     & \multirow{2}{*}{Insurance options}                                                               & Thornburg Mortgage                              \\ \cline{3-3} 
                                                                                     &                                                                                                  & TD Banknorth                                    \\ \cline{2-3} 
                                                                                     & Equipment for survival                                                                           & Orvis                                           \\ \hline
\multirow{7}{*}{\textbf{\begin{tabular}[c]{@{}c@{}}OTHER\\ PERCEPTION\end{tabular}}} & Islam in general                                                                                 & Moinuddin Chishti                               \\ \cline{2-3} 
                                                                                     & \multirow{4}{*}{Islam in Western countries}                                                      & Islam in Poland                                 \\ \cline{3-3} 
                                                                                     &                                                                                                  & List of mosques in the United Kingdom           \\ \cline{3-3} 
                                                                                     &                                                                                                  & Ed Husain                                       \\ \cline{3-3} 
                                                                                     &                                                                                                  & Bhagwan Shree Rajneesh                          \\ \cline{2-3} 
                                                                                     & Islam in non-Western countries                                                                   & Islam in Qatar                                  \\ \cline{2-3} 
                                                                                     & About Middle East                                                                                & Susa                                            \\ \hline
\multirow{15}{*}{\textbf{\begin{tabular}[c]{@{}c@{}}SELF\\ PERCEPTION\end{tabular}}} & National identity                                                                                & Juan Carlos I                                   \\ \cline{2-3} 
                                                                                     & Religious identity                                                                               & Church of the Holy Sepulchre                    \\ \cline{2-3} 
                                                                                     & \multirow{2}{*}{Activism in general}                                                             & Kyle Kulinski                                   \\ \cline{3-3} 
                                                                                     &                                                                                                  & Andrew O'Hagan                                  \\ \cline{2-3} 
                                                                                     & \multirow{5}{*}{Anti-Islam activism}                                                             & James Wickstrom                                 \\ \cline{3-3} 
                                                                                     &                                                                                                  & Quilliam (think tank)                           \\ \cline{3-3} 
                                                                                     &                                                                                                  & Tommy Robinson (activist)                       \\ \cline{3-3} 
                                                                                     &                                                                                                  & Sebastian Gorka                                 \\ \cline{3-3} 
                                                                                     &                                                                                                  & Paul Weston (politician)                        \\ \cline{2-3} 
                                                                                     & \multirow{6}{*}{\begin{tabular}[c]{@{}c@{}}Violent history of \\ Western countries\end{tabular}} & Grdelica train bombing                          \\ \cline{3-3} 
                                                                                     &                                                                                                  & Greek War of Independence                       \\ \cline{3-3} 
                                                                                     &                                                                                                  & Legitimacy of the NATO bombing of Yugoslavia    \\ \cline{3-3} 
                                                                                     &                                                                                                  & Charles B. McVay III                            \\ \cline{3-3} 
                                                                                     &                                                                                                  & The Man with the Iron Heart                     \\ \cline{3-3} 
                                                                                     &                                                                                                  & List of events named massacres                  \\ \hline
\end{tabular}
\caption*{Organization of the 69 articles that survived all filters (right column, same as Table~\ref{tab-all-reacted}) into Specific categories (middle column) and broad categories (left column).}
\end{table}


\clearpage
\section{Dynamics of total visits to English Wikipedia in relation to the attacks} \label{appendix:c}

\begin{figure}[!hbt]
\centering
\includegraphics[width=\textwidth]{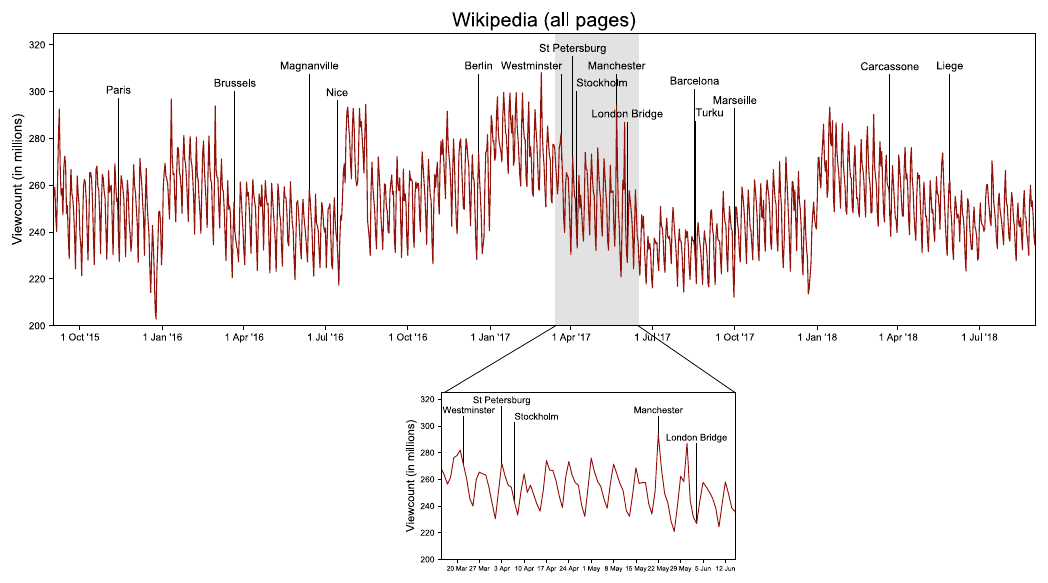}
\caption{The time series of daily viewcounts to all articles in English Wikipedia excluding only stubs and redirects. Top panel: the time series with dates of the attacks shown. Bottom panel: the same time series for 3 months in 2017 during which five attacks occurred (dates shown on $x$-axis are all Mondays).}
\label{fig-total-viewcounts}
\end{figure}

Do terrorist attacks divert attention from other simultaneously ongoing concerns? Here, we examine the time series of daily viewcounts to all articles cumulatively, excluding only stubs and redirects (see section \ref{Exclusion}, immediately after the exclusion Step 1). The obtained time series is shown in Fig.~\ref{fig-total-viewcounts} (top panel), in which we marked our sample of attacks. For better clarity, in the bottom panel of Fig.~\ref{fig-total-viewcounts} we zoom to a 3-month period in 2017 during which five attacks occurred.  \\[0.2cm]
The total views vary from 200 to 300 million per day. There are pronounced and stable weekly oscillations, with more activity during weekdays and less during the weekend. There are less pronounced seasonal oscillations, with slightly more activity during the early months of the year (winter in the northern hemisphere). As the Internet becomes ever more accessible, we expected to find more viewing activity in 2018 compared to 2015. Surprisingly, we found none. \\[0.2cm]
None of the considered terrorist events seems to affect the overall viewing dynamics, as clear from the bottom panel of Fig.~\ref{fig-total-viewcounts}. Weekly oscillations remain statistically stable immediately after any of the attacks. In the eyes of the global English-speaking public, terrorism is only one of very many concerns ongoing in parallel. Attacks might have diverted some attention from routine concerns, but the intensity of this diversion is too small to be recognized against the background of normal fluctuations. 


\clearpage
\section{Bench-marking against reactions to other events} \label{appendix:d}

What else, besides terrorist attacks, diverts public attention to these issues? We investigate this by systematically computing Excess Attention $Z$ for all articles from Table~\ref{tab-all-reacted}, starting with the Paris 2015 event. For each article, we isolate the reactions with $Z>3$ that happened at least two weeks away from any of the attacks. This led to a list of non-terrorism-related reactions for each article. \\[0.2cm]
We looked for patterns in this list but found none to be very clear. For a typical article, we found about 10--30 reactions to events other than terrorism. This number seems relatively constant across all broad categories. Overall, articles with more general content (e.g. \textit{Terrorism in Europe}) had more, and articles with more focused contents (e.g. \textit{Terrorism in Greece}) had less non-terrorism-related reactions. While broad categories react (in specific ways) to acts of terrorism, they also react to other events, albeit in a less consistent manner. This comes across as natural, since terrorism is not the only public concern. For instance, we found the viewcount of several articles from Table~\ref{tab-all-reacted} to spike on 25 January 2017. Yet many events could have been responsible: the inauguration of US President Donald Trump, withdrawal of the US from the Trans-Pacific Partnership, Oscar nominations, Syrian civil war, etc.  \\[0.2cm]
Along the same lines, we checked what else draws attention to the 15 articles devoted to the attacks themselves. To this end, we looked for large $Z$ away from all attacks. We found a few non-terrorism-related reactions, but much less than for articles from Table~\ref{tab-all-reacted}. It appears that the best `markers' of terrorist attacks are articles devoted to the attacks themselves. At any rate, the message of the above analysis is that too many events occur in the world simultaneously for any clear-cut result on non-terrorism-related reactions.


\clearpage
\section{Robustness checks} \label{appendix:e}

Critical for reaching our results was the choice of exclusion criteria. Indeed, many of our choices may appear arbitrary, leading one to think that different exclusion choices could have led to different results. For this reason, we checked whenever possible the robustness of the results against variations of the exclusion criteria. Upon very close examination we found no significant impact of such variations on our results, i.e., the overall picture (such as the structure of broad categories) still holds even with non-trivial modifications of our exclusion choices. \\[0.2cm] 
In the same context, recall that in defining Excess Attention, we implicitly assumed that daily viewcounts (in fact, their weekly averages) are normally distributed around the mean, which may not always be the case. But in practice, as clear from Fig.~\ref{fig-four-example-time-series}, $Z>3$ is a sufficient guarantee for a redirection of attention, even if viewcounts are not (perfectly) normally distributed. Upon excluding articles with $Z<3$, the only false positives we found (and examined manually, see \ref{Exclusion}) were the issues that accidentally coincided with the attacks. This confirms the validity of our $Z>3$ criterion regardless of viewcount distribution. \\[0.2cm]
On the other hand, our results could benefit from an independent verification via sources other than Wikipedia. These would rely primarily on social media, which nowadays are a standard channel of communication. This is an important direction for future work since it can elucidate the public reaction from a wider perspective. Still, each tweet is just one particular message (usually) designed by one person. Even if this tweet is very popular, its popularity might be due to a lack of a systematic r\'epertoire of tweets from which the audience could choose the best ones. Alternatively, one could verify our results via surveys and interviews: Ask people directly what interests them after a terrorist attack. However, there is a clear difference between what people do and what they report. Surveys typically measure the reported behavior only. In contrast, our insights are in the actual views of Wikipedia, hence revealing the \textit{real behavior}.

\end{appendices}


\printbibliography


\end{document}